# An Update on the Student Exoplanet Programme

Banks, T*, Rhodes, M.D.,^ & Budding, E.#

*: Data Science, Nielsen, 200 West Jackson Blvd, Chicago, IL 60606, USA; Physics & Astronomy, Harper College, 1200 W Algonquin Rd, Palatine, IL, 60067, USA

^: Brigham Young University, Provo, UT 84602, USA

#: Dept. of Physics, Canakkale Onsekiz Mart University, TR17020, Canakkale, Turkey; Dept. Physics & Astronomy, University of Canterbury, New Zealand;  SCPS, Victoria University of Wellington, P.O. Box 600, Wellington, New Zealand; Carter Observatory, 40 Salamanca Rd, Wellington, New Zealand

**Abstract:**

An update is given on the exoplanet research collaboration between Nielsen (a marketing research company), Brigham Young University, and NZ universities with the National University of Singapore, which has been expanded to include a community college in the US. Key achievements from the past year are outlined, including density estimates for HD 209458 and Kepler-1 from radial velocity and transit fits. A comparison between the "WinFitter" optimizer and other techniques is outlined, showing that WinFitter estimated statistical errors are essentially in line (bar a scaling proportion) with those estimated via Markov Chain Monte Carlo techniques. This work was completed by undergraduates at both teaching and research universities, who built the modeling code from first principles. This reinforces the paper's recommendation that such projects are excellent introductions to science and astronomy.

## (1) Introduction:

For centuries humanity had speculated about the existence of other planets orbiting other stars to our own.  25 years ago the discovery of a planet orbiting the solar-like star 51 Peg (Mayor & Queloz, 1995) opened the floodgates, leading to a vigorous expansion in the number of known exoplanets.

51 Peg was the first widely accepted exoplanet discovered orbiting a solar-like star, using radial velocity measurements just as Struve (1952) had suggested could result in such a discovery of a massive planet close to its host: `A planet ten times the mass of Jupiter would be easy to detect, since it would cause the observed radial velocity of the star to oscillate with ± 2 km per second'.  It is worth noting that 51 Peg was not the first exoplanet to be found.  Radio pulsar timings by Wolszczan & Frail (1992) had provided convincing evidence to planetary masses outside our solar system. The radial velocity work of Latham *et al.* (1989) and Hatzes & Cochran (1993) had provided evidence suggesting objects of planetary masses orbiting main sequence stars. Similarly, Campbell *et al.* (1988) proposed a planetary candidate in a 2.7 year orbit in the binary Gamma Cephei system, based on a radial velocity curve of 25 metre per second amplitude. However Irwin *et al.* (1989) and Walker *et al.* (1989) later cast doubt on the interpretation, noting that their velocity standards showed roughly annual variations that correlated with their chromospheric activity. Walker *et al.* (1992) refined the period to 2.52 years noting that "an explanation… in terms of … a Jupiter-mass planet in a high circular orbit …is still viable" but expressed concern that such a massive planet could truly exist so close to its host star. Indeed, the strange nature of 51 Peg's planet (an orbital radius some 5% that of Earth's, a 4.23 day orbit, and a mass similar to Jupiter's) raised eyebrows at the time of its discovery announcement. The field of planet discoveries had had many false alarms in the past, such as Peter van de Kemp's claims that unfortunately turned out to be instrumental in cause (see Wenz, 2019). Nevertheless, following the 51 Peg announcement further work by the Europeans and a US team headed by Marcy & Butler quickly led to further exoplanet discoveries, followed shortly by the discovery of what could only be planetary transits in HD 209458 (see Sullivan & Sullivan, 2003, to learn about the New Zealand connection to this work). As evidence of the rapid expansion of the field, as of May 2020 the NASA Exoplanet Archive (NEA) listed 4,154 confirmed exoplanets. The importance of 51 Peg is that it signaled the "beginning of the flood", the start of the results from established search campaigns and the beginning of the challenge to our understanding of how solar systems form and evolve.

**(2) Student Study of Exoplanets:**

The field of exoplanets is clearly a highly active one. As Banks & Budding (2018, 2019) outlined in this journal, the field is one where undergraduate students can make active research contributions. Transit and radial velocity data are freely available via the NEA, frequently providing high quality data from large terrestrial telescopes or space-based missions such as Kepler (see Borucki *et al.*, 2010) and TESS (see Ricker et al., 2015). Students can analyse real data and draw real conclusions, which if done well can lead to publications in research journals. Such articles can be used in support of graduate school applications, as indeed one of the former students in the programme successfully did for entry to the PhD programme at Duke University.

This note is an update on the programme, outling activity in the last year. The programme had concentrated on final year statistics undergraduate projects at the National University of Singapore (NUS), and this continued in 2019 with Eden Choo Jing Yu (see Figure 1) undertaking an exoplanet project. Two other honours student projects were also supervised by some of the authors, but these projects are not discussed here as they focussed on the application of advanced analytics to human resource matters.

With the relocation of one of the authors to Chicago, we have been able to widen the participation to include final year astronomy students at William Rainey Harper College, a public community college opened in 1967 (see Figures 2 and 3). Community colleges are primarily two year public tertiary institutions in the US, typically supported by local tax revenue (in this case, the village of Palatine). They grant certificates, diplomas and associate degrees. Many Harper students transfer after graduation to a four year university to complete a bachelors degree (crediting across their two community college years, and only needing to do an additional two years to reach their Bachelors degree --- something that normally takes four years to complete) and later progress to graduate study should they wish.

Harper has an active astronomy teaching programme, including the use of its Karl G. Henize observatory which is equipped with a 14 inch Schimdt-Cassegrain telescope. The astronomy department invited the authors to pilot a semester long "independent study course" on exoplanets. One of our goals was to teach the students how to use the R statistical programming language and its associated tools (Jupyter and RStudio), practical skills and experience valuable in the field of data science. R was chosen in preferance to Python and Julia for its strong integration with statistical techniques. The students built from first principles models for radial velocities (using equations from Haswell, 2010) and transit light curves (using equations from Mandel & Agol, 2002), and then applied simple optimisation techniques to derive estimates for the masses, orbital inclination and radius, planetary radius, and stellar radius. Where they had both estimates of the mass and radius of a planet, they could calculate the density, leading to discussion about the nature of the planet. The students also learnt how to download and prepare for analysis data from the NEA, both radial velocities and transit information.

We had originally planned for the Harper students to prepare a poster presentation as one of their deliverables, to be presented at the annual "Astronomy Day" hosted by Harper and a local amateur astronomy association. We had also hoped to have the students present their posters at a meeting of the American Astronomical Society (AAS), which was being held in the town of Madison (Wisconsin), only a couple of hours drive from the college. Unfortunately the COVID-19 pandemic led to not only the cancellation of the annual event and the AAS meeting switching from physical to virtual, but to a change from face-to-face presentations and classes to WebEx video conferencing sessions. Indeed, the entire campus was closed for both students and faculty during the closing months of the spring semester. This certainly made both teaching and learning harder, but we all adapted! The final deliverable for the students therefore reduced to a written report, of about a dozen pages in length and following the layout of a scientific report.

The pandemic had a lesser negative impact for the NUS students as we had been using teleconferencing for the last few years already, given the international locations of their supervisors and examiners. The deliverable for the NUS students is a reseach thesis, typically of length 100 pages, together with a 45 minute presentation followed by an oral defense involving both internal and external examiners. With the increase in the number of honour students at NUS increasing dramatically the supervision load on Faculty, only the top students in this group are allowed to pursue individual theses with the other students working on group projects. The NUS students also built their models from first principles, but as statistics students were expected to take their analyses further, such as applying deep learning or advanced Bayesian techniques. Some of these topics have been discussed in Banks & Budding (2018, 2019) and are therefore not repeated here.

**(3) Results**

**(3.1) Singapore:**

Eden's project focussed on the accuracy of results (such as the orbital distance, orbital inclination, planetary radius, etc.) from various optimisation techniques applied to light curves, i.e., he focussed on understanding transits and not the radial velocity data. A transit is detected when an observed star dims, the planet orbiting it passes through our line of sight and obscures part of the stellar surface from our view. Careful modelling and analysis can tell us about the size of the planet and the orientation of its orbit.

In particular he was interested in comparing the estimates from WinFitter (software developed by Drs. Budding and Rhodes for modelling eclipsing binaries, see Rhodes & Budding, 2019) with those from other techniques. Winfitter is written in Fortran-95, giving it high computational speed. Eden had to integrate the relevant Fortran functions and subroutines with R, allowing him to use the sophisticated Winfitter model with optimisation techniques available inside R. These techniques included the Levenberg-Maarquadt (LM), boot strapping, Metropois-Hastings Monte-Carlo, and the No U-Turn Monte Carlo (NUTS, see Hoffman & Gelman, 2011) methods. The last two techniques are Markov Chain Monte Carlo (MCMC) methods. MCMC is a stochastic procedure that repeatedly generates random samples that characterize the distribution of parameters of interest, or to put it another way, a method that draws samples randomly to approximate the probability distribution of parameters in a model. MCMC allows us to better understand the "accuracy" of parameter estimates. However this can take considerable time, being basically a "drunkard's walk". It is normal to run many thousands of steps, building up a "chain" of estimates from which the distributions can be estimated. The complete WinFitter program uses an algebraic approach to estimate uncertainties, taking considerably less time. Eden was interested to see if the (complete program) WinFitter estimates were similar to those from the other techniques, and in particular if there was a simple relationship between estimates from the different techniques (such as a scalar).

To summarise, Eden compared several techniques. The first was the complete WinFitter program --- which was both the WinFitter model and its own optimisation technique. He used the WinFitter model with other optimisation techniques, such as boot strapping, LM, and Metroplis

Hasting. Eden compared the statistical error estimates from these methods with those from the complete WinFitter program.

Eden modelled 8 systems --- Kepler-2b, -5b, -77b, -428b, -491b, -699b, -706b, and GJ 357b. Data for the last system came from the TESS mission and for all the others from the primary Kepler mission. He found that his point estimates for parameters such as the inclination and radii were in good agreement with the literature, boot stapping led to surprisingly low estimates of the uncertainty, he considered these unrealistic. He then moved to the Metropolis-Hastings method and wrapped that around the WinFitter model, this had good convergence for all systems but GJ 357b. This final system was a challenging one, being a suspected "super-Earth" (approximately 20% larger in radii than the Earth) with a correspondingly smaller impact on the light curve. Despite running 50,000 steps, his solution was not yet stable for this planet.

As a cross-check to the optimisation methods using the WinFitter model (but not its built-in optimisation technique), Eden also implemented the Mandel & Agol equations, intergrating this with the NUTS method. NUTS was available inside the STAN programming language and not R. Unfortunately STAN cannot be easily integrated with Fortran code, so Eden was not able to combined NUTS and the WinFitter model. Eden found good agreement between the point estimates from Winfitter and the NUTS code but he did not find perfect agreement between the two methods in their estimation of accuracy. There did appear to be linear trends between the two, for instance in that WinFitter's statistical error estimates for the planet radius were linearly less than those from the NUTS and Metropolis-Hastings runs. Inclination uncertainties were in good agreement, while stellar radii statistical errors appeared to be systematically less in WinFitter estimates.

Further work is needed. More systems to be modelled to firm up the relationships between the different optimisation techniques, building off these tantalising results. We also need to convert the WinFitter model into STAN/R code so that the NUTS methodology can be used. Then we will be "comparing apples with apples". If WinFitter error estimates can definitively be shown to be similar to those from MCMC methods (after scaling, for example), its lower computational footprint could make the program appealing for large scale modelling of databases searching for planets. Given the high number of discovered planets, the field is moving towards population studies. Realistic and reliable estimates of parameter estimates will be critical in such work. Even a cursory examination of the summary pages of the NEA show clearly the wide range of statistical error estimates for a given system, which can be orders of ten in size.

### (3.2) Chicago

Two students took part in the pilot of the independent research study course on exoplanets, Faraz Uddin and Fernando Flores (see Figure 4). The course started with an overview on exoplanets before moving into the physics behind the Doppler and Transit models together with an introduction to programming in R (neither of the students knew this language). By the end of the course, the students had written R code that used optimisation to arrive at best fit parameters to radial velocity and transit data, generally sourced from the NEA. The models were for circular orbits (keeping the models easier to implement) and the simplest model of limb darkening was used (linear limb darkening). Both students tested their radial velocity code against data for 51 Peg --- for example, Fernando derived a mass of 139 times that of Earth, in reasonable agreement with the 146 of Martins *et al.* (2015) and the 148 of Mayor & Queloz (1995). This gave the students confidence that their code was operating correctly.

Fernando then moved on to try to estimate the density of HD 209458, using photometry from the Hubble Space Telescope (Brown *et al.,* 2001) and radial velocities from the 10-m Keck (Butler *et al.,* 2006). He first fitted (see Figure 5) the transit data, and used the resulting inclination in his radial velocity modelling, finding a mean density of 330 kg per cubic metre. This is in good agreement with the value estimated by del Burgo & Allende Preietoz (2016), being within 3% of their value. Francisco correctly concluded he was looking at a gaseous planet. The optimisation routines we used only calculated point estimates, so we moved on to a methodology allowing this. An MCMC analysis was made for the transit data using R/STAN, but we ran out of time to build the code using the same technique on the radial velocity data. We therefore ran EXOFAST (Eastman *et al.*, 2013) which fitted both the transit and radial velocity data together, resulting in an estimated density of 341 kg per cubic metre and an equilibirum temperature of 1453 Kelvin. The mass ratio was 0.000575, inclination 86.63 degrees, and the quadratic limb darkening coefficients were estimates as 0.32 and 0.28 respectively. A low orbital eccentricity of 0.033 was derived, together with a semin-major axis of only 0.0475 AU. We followed up with an MCMC fit using EXOFAST, which gave the following results:
- Inclination: 86.62 degrees, +0.32, -0.52
- Semi-major axis in stellar radii: 8.74, +0.38, -0.59
- Linear limb darkening coefficient: 0.29, +/- 0.03
- Quadratic limb darkening coefficient: 0.35, +/- 0.04
- Planet mass: 0.778 Jupiters, +0.089, -0.088
- Planet radius: 1.467 Jupiters, +0.116, -0.075
- Planet density: 303 kg per cubic metre, +/- 6
- Equilibrium temperature: 1675 Kelvin, +62, -42
- Mass ratio: 0.00053, +/- 0.00006
- Mean orbital radius: 0.0507 AU, +/- 0.0008
- Eccentricity: 0.049,+ 0.061, -0.034

It is interesting that some of these estimates differ from the simple point estimates made by EXOFAST. Further work is needed to investigate, and will be made in the next iteration of the Harper College course. However it worth noting that the results are in good agreement with those listed by the NEA for this system, bar a few such as the effective temperature which is a few hundred Kelvin too high (the point estimate gave a value in better agreement with the literature). The MCMC fit still shows an eccentricity, which the authors believe is probably over-fitting the data.

Faraz followed a similar path, examining in depth Kepler-1 using radial velocity data from the SDSS III Multi-object APO Radial Velocity Exoplanet Large-area Survey (MARVEL; Thomas *et al.,* 2016) and (short cadence) photometry from the Kepler mission. The radial velocity data were relatively noisy and sparse (being 19 data points, see Figure 6).  The optimisation routines consistently settled on a good fit to the data, leading to a planet mass equivalent to 398 Earths.  This is in good agreement with the literature value of around 380 Earths.  If we had had more time we would have tried to source additional data, the papers we consulted such as Winn *et al*. (2008) concentrated on limited phases but if carefully combined with the Thomas data set these might have helped constraint the fit further. We continued with the data set on its own as it led to interesting class discussions about the difficulty of analysis and how to critically interpret results and claims[1] by other researchers.  The transit fit (see Figure 7) was closer to the literature, coming in at a value of 13.4 Earth radii (which can be compared, for example, with the value of 14.186 from Berger *et al.,* 2016, at the high end and 13.02 (Barclay *et al.,* 2012) at the low end of the literature).  This leads to a mean density of 650 kg per cubic metres.  The value is lower than the literature mean of 800 to 900 kg per cubic metres. A formal statistical error analysis would be needed to see if this difference is significant (a start was made with the transit fit, as shown in Figure 8, but needs to be extended to the radial velocities).  However it confirms a gaseous mean density.  We did attempt to run EXOFAST on the system, but the MCMC did not converge.  Other studies such as Yi *et al.* (2017) had noticed similar convergence issues with this system.  Our experience with EXOFAST is increasing with additional systems being modelled, such as Kepler-12b, -14b, and -15b by Shi Yuan Ng (a previous NUS student, whose work was outlined in Banks & Budding, 2019), whose work is in final preparation for publication.  We will revisit using EXOFAST with Kepler-1 in the next iteration of the course.

**(4) Discussion:**

Our conclusions are that both these programmes were successful.  Not only did the students come up with interesting research results, they were able to build models from first principles, implementing them in a popular statistical programming language and learning good programming practices (such as documentation, via the Jupyter notebook).  These are marketable skills and valuable in the commercial world.  We have outlined other benefits in Banks & Budding (2018, 2019). We look forward to both programmes continuing.

It is certainly hard work being the supervisor for such projects, but we believe it is worthwhile.  Frequent communication with the students is key to help them progress. The rewards include seeing the increased confidence of the students as they master the research, imparting useful work skills that will help their careers, and increasing their scientific literacy. One student commented that he will be better able to critique and assess claims in the media, for instance around "habitable" planets. The students also provide us with the extra capacity to assist with our research efforts.

---

[1] After all, how often do we hear about "habitable" or "earth-like" planets in the press? The differences led to class discussion about the scientific method, the importance of data, the habitable zone and what it means, and the need for scientific literacy and critical interpretation.

We believe so much in these rewards that we have expanded our efforts, as outlined in this note. In particular we were pleased to see how well the Harper students took to research, especially given that community colleges are essentially teaching universities. The success of the Harper pilot will lead to further research-based study courses, for instance we are currently looking into an expansion for the Fall semester to cover variable stars.

We hope that this note will encourage others to try similar programmes in New Zealand. Such "citizen science" could help increase not only the number of active members of astronomical societies, but lead to more professional astronomers in the future. We believe that similar programmes could be possible for final year secondary school students. In this case, we would not recommend optimization techniques but simple graphical methods to read off estimates of the planet's radius, for example. "Bite" size projects might be preferable to the longer duration projects described here. While we have concentrated on exoplanets, primarily due to the easy availability of high quality date and the popularity of the topic with the general public, there are many suitable subfields of astronomy such as lunar occultations, variable stars, and even photometry of star clusters. Perhaps local amateur astronomers could partner with schools to provide access to telescopes and data, working with teachers and educational professionals to build an impactful learning path that supports the teacher's wider learning objectives through the use of an exciting field? The key is to work with learning professionals and to link with the learning goals and objectives they wish to meet. Teaching at tertiary educational intuitions has fewer restrictions than teaching at the secondary level, especially when external targets (such as national examinations) need to be kept in mind and learning is more towards specific goals. The authors do not have direct experience building secondary level projects, and so will not comment further other than to say we look forward to hearing and learning from others about their efforts.

The authors had the pleasure to work with (e.g., Budding *et al.*, 2004), or in the case of TB study with, Professor Denis Sullivan of Victoria University of Wellington. Denis firmly believed in the importance of teaching physics and astronomy, carefully crafting his teaching materials. He was instrumental in not only supervising research students, but also setting up and running a very popular introductory astronomy course. He believed it was important to bring the wonder of the universe to students, both to enrich their appreciation of it and to build support for the science. As Denis used to say: "the more you put in, the more you get back". We agree with his assertion and how this relates to the building up of the next generation of astronomers. We hope that he would have been pleased with this initiative, and how we took his teaching to heart.

**(5) Acknowledgments:**

This paper includes data collected by the Kepler mission and obtained from the MAST data archive at the Space Telescope Science Institute (STScI). Funding for the Kepler mission is provided by the NASA Science Mission Directorate. STScI is operated by the Association of Universities for Research in Astronomy, Inc., under NASA contract NAS 5–26555. This research has made use of the NASA Exoplanet Archive, which is operated by the California Institute of Technology, under contract with the National Aeronautics and Space Administration under the Exoplanet Exploration Program. This paper makes use of EXOFAST (Eastman *et al.,* 2013) as provided by the NASA Exoplanet Archive. Additional help and encouragement for this work has come from the National University of Singapore, particularly through Prof. Lim Tiong Wee of the Department of Statistics and Applied Probability. TB gratefully acknowledges the supervision of the late Prof. Denis Sullivan for his Masters and PhD (together with Drs. Edwin Budding and Richard Dodd), and for teaching the importance of teaching. The authors extend their sincere condolences to Denis' family.

**References:**

- Banks, T., & Budding, E., 2018, Southern Stars, 57(2), 10
- Banks, T., & Budding, E., 2019, Southern Stars, 58(2), 14
- Barclay, T., Huber, D., Rowe, J.F., et al., 2012, ApJ, 76, 53
- Berger, T.A., Huber, D., Gaidos, E., & van Saders, J.L., 2018, ApJ, 866, 99
- Borucki, W. J., Koch, D., Basri, G., Batalha, N., et al., 2010, Science, 327, 977
- Brown, T.M., Charbonneau, D., Gilliland, R.L., Boyes, R.W., & Burrows, A., 2001, ApJ, 552, 699
- Budding, E., Sullivan, D., & Rhodes, M., 2004, in *Transits of Venus: New Views of the Solar System and Galaxy*, Proc. IAU Colloq. No. 194, 386
- Butler, R.P., Wright, J.T., Marcy, G.W., Fischer, D.A., Vogt, S.S., Tiney, C.G., Jones, H.R.A., Carter, B.D., Johnson, J.A., McCarthy, C., & Penny, A.J., 2006, ApJ, 646(1), 505
- Campbell, D., Walker, G. A. H., & Yang, S., 1988, ApJ, 331, 902
- del Burgo, C., & Allende Preito, C., 2016, MNRAS, 463, 1400
- Eastman, J., Gaudi, BS., & Agol, E., 2013, PASP, 125, 83
- Hoffman, M.D., & Gelman, A., 2014, Journal of Machine Learning Research, 15, 1351
- Haswell, C., 2010, "Transiting Exoplanets", Cambridge University Press
- Hatzes, A. P., & Cochran, 1993, ApJ, 413, 339
- Irwin, A. W., Campbell, B., Morbey, C. L., Walker, G. A. H., & Yang, S., 1989, PASP, 101, 147
- Latham, D. W., Mazeh, T., Stefanik, R. P., Mayor, M., & Burki, G., 1989, Nature, 339, 38
- Mandel, K., & Agol, E., 2002, ApJ Letters, 580(2), 171

- Martins, J.H.C., Santos, N.C., Figueira, P., et al., 2015, A&A, 576, A134, 9
- Mayor, M., & Queloz, D., 1995, Nature, 378, 355
- Rhodes, M.D., & Budding, E. 2019, "WinFitter User's Manual", https://michaelrhodesbyu.weebly.com/astronomy.html
- Ricker, G. R., Winn, J. N., Vanderspek, R. et al., 2015, Journal of Astronomical Telescopes, Instruments, and Systems, Volume 1, id. 014003
- Struve, O., 1952, The Observatory, 72, 199
- Sullivan, D.J., & Sullivan, T., 2003, Baltic Astronomy, 12, 145
- Thomas, N., Ge, J., Grieves, N., Li, R., & Sithajan, S., 2016, PASP, 128, 45003
- Walker, G.A.H., Yang, S., Campbell, B., & Irwin, A.W., 1989, ApJ, 343, L21
- Walker, A. H. W., Bohlender, D. A., Walker, A. R., Irwin, A. W., Yang, S., & Larson, A., 1992, ApJ, 396, L91
- Wenz, J., 2019, "The Lost Planets", MIT Press
- Winn, J.N., Johnson, J.A., Narita, N., Suto, Y., Turner, E.L., Fischer, D.A., Butler, R.P., Vogt, S.S., & O'Donovan, F.T., 2008, ApJ, 682(2), 1283
- Yi, J., Banks, T., Budding, E., & Rhodes, M., 2017, ApSS, 362, 112

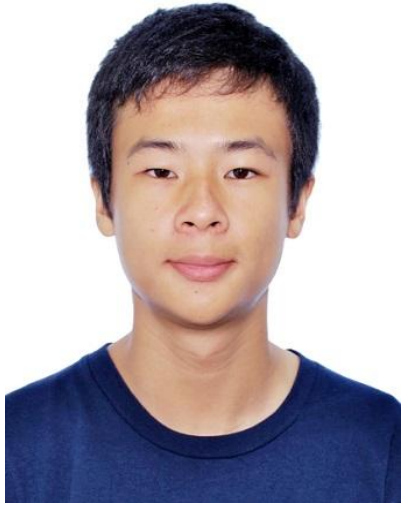

**Figure 1:** Eden Choo, an Honours student at National University of Singapore (photo supplied by Eden).

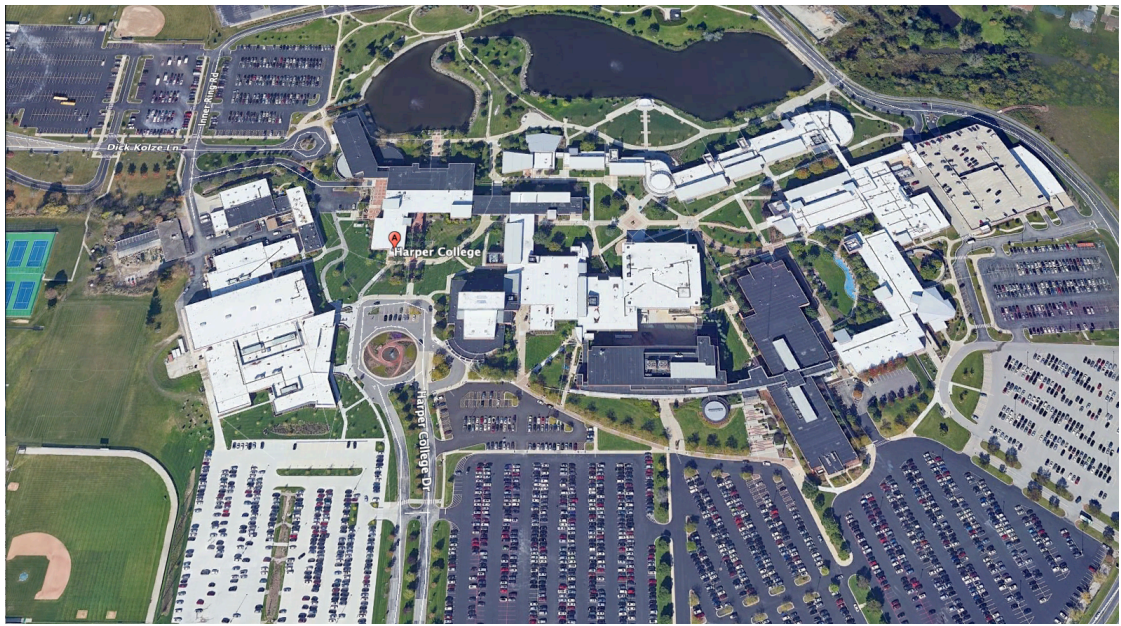

**Figure 2:** an aerial view of the part of the main (200 acre) campus of Harper College, which hosts most of the more than 15,000 enrolled students. The astronomy group is housed in the building in the centre of the image, just above the carparks. The teaching observatory is located at the top of the image, above the lakes and in the grassy area near the ring road (image from Google Earth).

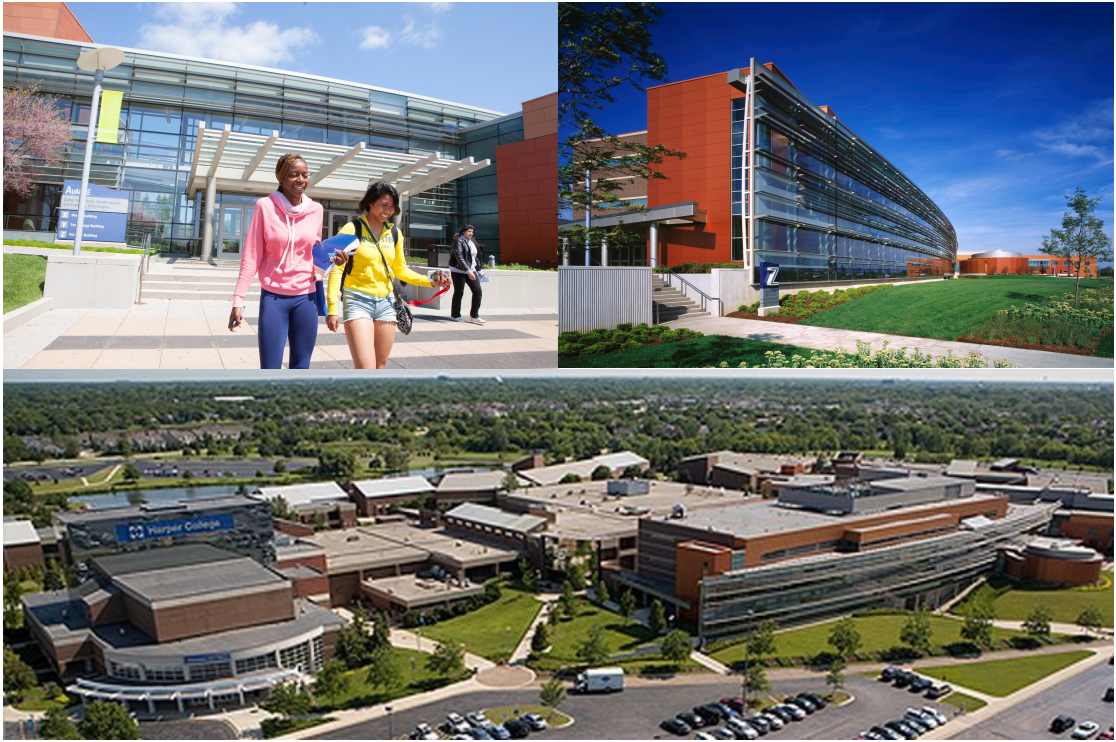

**Figure 3:** Photos of Harper College showing the entrance to the science building astronomy is hosted in, a view of that building to the side of main entrance, and an aerial view showing the location of the building in the campus (photos courtesy of Harper College).

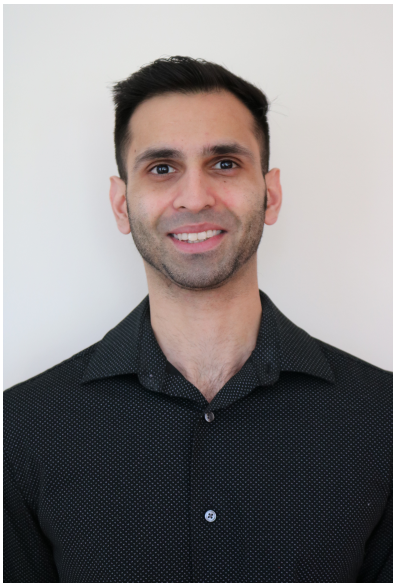

Figure 4: Photo of Faraz Uddin, research student at Harper College.

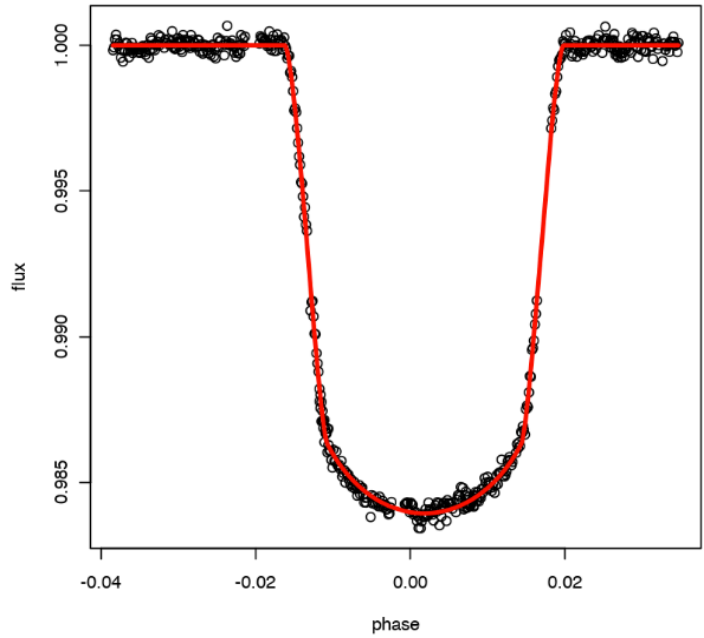

**Figure 5:** Fernando's model fit (the red line) to HST data for HD 209458. Phase is in radians, and (out of transit) flux normalised.

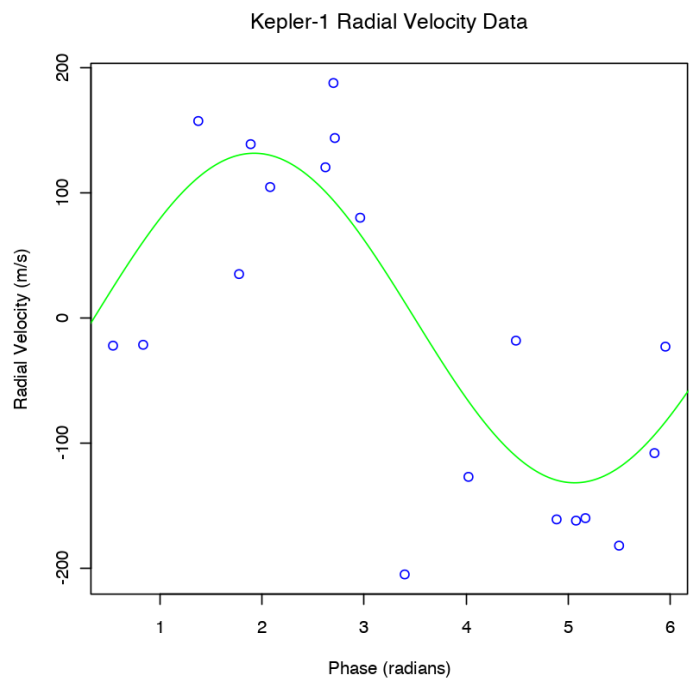

**Figure 6**: Faraz's radial velocity fit for Kepler-1 b. The green line is his optimized model fit.

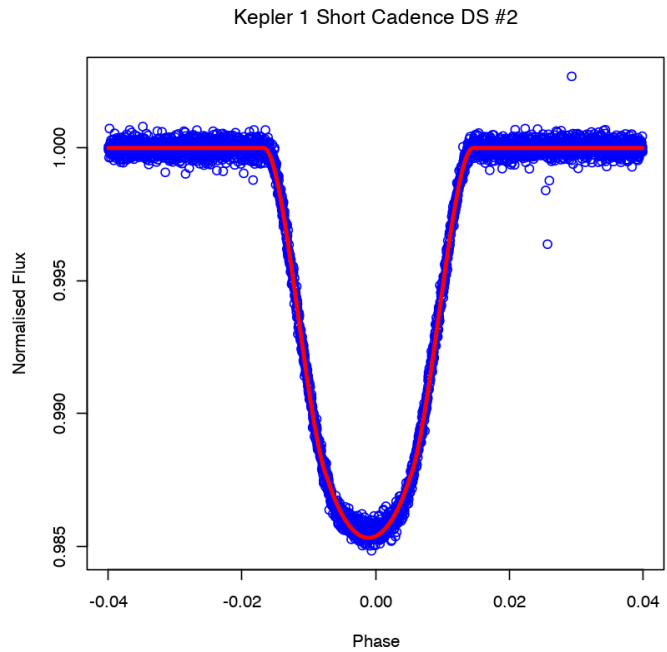

**Figure 7:** Faraz's fit to transit data from Kepler's Quarter 2 (first data set of the three available) short cadence data. The data were not binned. The red line is his model fit. Phase is in radians.

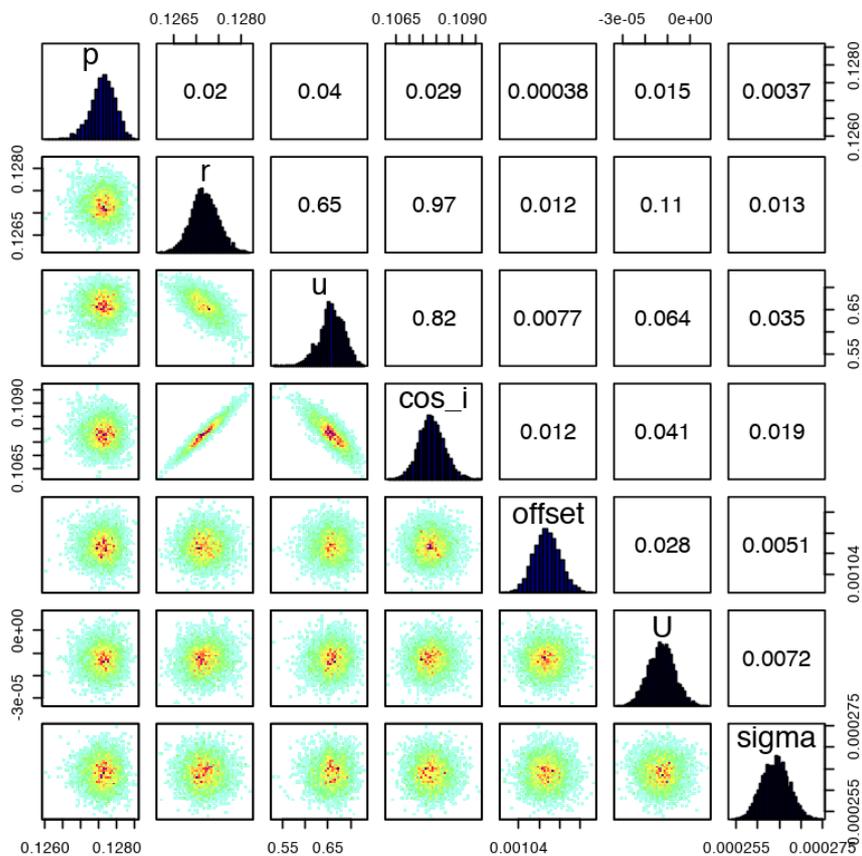

**Figure 8:** MCMC results to the Kepler 1b transit data. As the data are quite clean, only 10,000 steps (points) were required across 4-chains and are plotted here. The density maps can be used to estimate statistical 'errors' (or uncertainties) in the optimized parameters. The more "red" the subplot, the greater the number of steps falling into that region (and hence the more likely that solution). As might be expected there is a strong relationship between inclination (cos_i) and the planet radius (r). The limb darkening (u) also interacts with these terms, as could be expected. The remaining parameters are the stellar radius relative to the mean orbital radius (p), two adjustment parameters, and an estimate of the Gaussian noise in the data (sigma). A similar analysis would be needed for the radial velocity data and is something planned for the next iteration of the course at Harper College, now that we know that the students are able to reach this kind of analysis.